\documentclass[12pt]{iopart}

\usepackage{amsfonts}
\usepackage{amssymb}
\usepackage{graphicx}
\usepackage{bbold}
\usepackage{textcomp}
\usepackage{epsfig}
\usepackage{bm}
\usepackage{epstopdf}

\begin{document}
\title{On the Purity and Indistinguishability of Down-Converted Photons}
\author{C.~I.~Osorio, N.~Sangouard, and R.~T.~Thew}
\address{Group of Applied Physics, University of Geneva, 1211 Geneva 4, Switzerland}
\ead{robert.thew@unige.ch}
\begin{abstract}
Photons generated by spontaneous parametric down conversion (SPDC) are one of the most useful resources in quantum information science. Two of their most important characteristics are the purity and the indistinguishability, which determine just how useful they are as a resource. We show how these characteristics can both be accessed through Hong, Ou and Mandel (HOM) type interferences using a single pair source. We also provide simple and intuitive analytical formulas to extract their values from the depth of the resulting interference patterns. The validity of these expressions is demonstrated by a comparison with experimental results and numerical simulations. These results provide an essential tool for both engineering SPDC sources and characterizing the quantum states that they emit, which will play an increasingly important role in developing complex quantum photonic experiments.
\end{abstract}

The ability to engineer and characterize quantum photonic states is of paramount importance as the complexity of the experiments in quantum information science increases. In particular, it is fundamental to be able to combine multiple independent sources and this is usually done via interference, e.g. through a Bell state measurement (BSM).  At the heart of interference there are two concepts, two characteristics, of the photonic states: {\it Purity} - how close the real photons are to the theorist's ideal of a single mode quantum object, and the {\it Indistinguishability} - how alike are the different photons. Indistinguishability is a fairly clear concept and its role in interference experiments has been widely accepted for some time, in contrast to the purity, which is often ignored. 

The indistinguishability and purity can be increased either by filtering, or by engineering the source itself. As we will see, filtering is an option for some applications, but it is also a loss factor and potentially a significant one. If one wants efficient, even scalable, multi-photon systems, then one needs to go to the source. In this article we will focus on spontaneous parametric down conversion (SPDC) as this provides one of the most widespread and flexible resources for experiments in quantum information science. 

SPDC sources probabilistically generate two-photon states with diverse features in space and frequency and in those degrees of freedom the photons can be anywhere from maximally entangled to totally uncorrelated. A common way to characterize SPDC generated photons is by using two-photon interferometers. These were first introduced by Hong, Ou and Mandel \cite{Hong87} as a tool to measure the relatively narrow bandwidth of SPDC photons. The basic idea is to  observe the coincidence statistics at the output of a 50/50 beamsplitter as a function of the delay between two photons at the beamsplitter input. Due to their bosonic nature the photons may bunch, giving rise to the characteristic dip in the coincidences.   This interference effect will depend of the indistinguishability and purity of the photons and whether they are from the same, or independent, sources. Therefore, the question arises as to what information concerning the indistinguishability and purity can be extracted from these HOM-like interference patterns?

By focusing on the frequency degree of freedom, in this paper we discuss what information can be obtained from the interference patterns of different interferometers. We provide simple analytical formulas relating the SPDC parameters with the characteristics of the interference patterns. Such formulas can therefore be used to engineer photonic states with specific values of indistinguishability and purity. These are shown to be in good agreement with both experiment and numerical modeling. 

\section{The two-photon state}
In SPDC, pairs of photons are generated by the interaction of a classic pump beam and a nonlinear material. Conservation of energy and momentum set the relations between frequency and spatial distribution for the pump and the generated photons: signal and idler. The correlations between the degrees of freedom can be minimized by a careful design of the source \cite{Osorio08}, or can be erased by projecting the state into a single mode in one of the degrees of freedom.  In those cases,  the two-photon state is a product between a spatial and a frequency part. Here, we focus on the frequency part, but an analogous analysis is possible for the more general case.

When considering only the spectral component, the two-photon state generated in  SPDC  is described by the pure state $\rho=|\Psi\rangle\langle \Psi|$, where:

\begin{eqnarray}
\label{s2p}
|\Psi\rangle= \int d\Omega_s  d\Omega_i \Phi(\Omega_s,\Omega_i )\hat{a}^{\dagger}(\Omega_s)\hat{a}^{\dagger}(\Omega_i)  |\Omega_s, \Omega_i\rangle.
\end{eqnarray}

\noindent If we define $w_n$ as the frequency of the photons, then $\Omega_{n}=w_n-w_n^0$ gives the deviation from the central frequency $w_n^0$. All of the information about the frequency distribution of the generated pair of photons is described by the normalized mode function $\Phi(\Omega_s,\Omega_i )$. The state of a single photon is calculated by tracing out the other photon from the two-photon density matrix $\rho$, i.e.  $\rho_s=\mathrm{Tr}\left[|\Psi\rangle\langle \Psi|\right]_{i}$, which can be written as

\begin{eqnarray}
\label{dms2}
\rho_s= \int d\Omega_s  d\Omega_s'  \phi_s( \Omega_s, \Omega_s')  |\Omega_s\rangle \langle \Omega_s'|,
\end{eqnarray}
where the  single-photon mode function  is
\begin{eqnarray}
\label{mfs}
 \phi_s( \Omega_s, \Omega_s')= \int d\Omega_i  \Phi( \Omega_s,\Omega_i )\Phi^*( \Omega_s',\Omega_i ).
\end{eqnarray}

\noindent To explicitly calculate the mode function, consider a  Gaussian-pulsed laser with bandwidth $\sigma_p$, and a crystal with a length $L$, in that case $\Phi(\Omega_s,\Omega_i )$  is given by

\begin{eqnarray}
\label{mf}
\Phi(\Omega_s,\Omega_i )&=&N\exp{\left[-\frac{(\Omega_s+\Omega_i)^2}{4 \sigma_p^2}\right]}\mathrm{sinc}\left[\frac{\Delta_kL}{2}\right]\\\nonumber
&&\hspace{2.2cm} \times\exp{\left[ - i \Delta_k L  \right]}\exp{\left[-\frac{\Omega_s^2}{4 \sigma_s^2}-\frac{\Omega_i^2}{4 \sigma_i^2}\right]}.
\end{eqnarray}

\noindent Here  $N$ is a normalization constant, such that $|\Phi(\Omega_s,\Omega_i )|^2 = 1$ and the first exponential is due to the spectral distribution of the pump, assuming a Gaussian profile. The properties of the material appear through the phase-matching term $\Delta_k=k_p-k_s-k_i$ and the last exponential represents any possible spectral filtering of the photons, which we model as Gaussian functions with bandwidths $\sigma_s$ and $\sigma_i$.

Some approximations are useful when calculating the properties of the mode function in  \eref{mf}. For instance, by taking the first order terms in the Taylor expansion of $\Delta_k$, it is possible to write the phase-matching term as $\Delta_k=(N_p-N_s)\Omega_s+(N_p-N_i)\Omega_i$, where $N_{p}$, $N_{s}$ and $N_{i}$ are the inverse group velocities of the pump, signal and idler. Another common approximation is to replace the function  $\mathrm{sinc}(x)$ by $\exp{(-\alpha x^2)}$ where, if $\alpha=0.193$, as both functions have the same  FWHM. Under these approximations the mode function becomes, 

\begin{eqnarray}
\label{aproxxmf}
\Phi(\Omega_s,\Omega_i )&=& N  \exp{\left[-\frac{a}{4}\Omega_s^2-\frac{b}{4}\Omega_i^2-\frac{c}{2}\Omega_s\Omega_i -\frac{i}{2}(m\Omega_s+n\Omega_i)\right]}
\end{eqnarray}

\noindent where we define:

\begin{eqnarray}
\label{abc}
a&=&\alpha^2m^2 + 1/\sigma_p^2+1/\sigma_s^2;\\\nonumber
b&=&\alpha^2n^2 + 1/\sigma_p^2+1/\sigma_i^2;\\\nonumber
c&=&\alpha^2mn + 1/\sigma_p^2.\\\nonumber
\end{eqnarray}

\noindent The remaining factors $m=L(N_p-N_s)$ and $n=L(N_p-N_i)$, are dependent on the crystal length and the difference in the inverse group velocities for the different fields. 

 Writing the mode function as a product of exponentials as in \eref{aproxxmf} makes it possible to analytically calculate many quantities related to the two-photon state  \cite{Osorio08}. It also provides a useful and intuitive geometrical picture of what is happening in the frequency space of the joint function - the first three terms clearly describe an ellipse with the axes defined by $a$ and $b$ and the $c$ component defining a rotation about $\Omega_s$ and $\Omega_i$. The last term describes the relative phase between  $\Omega_s$ and $\Omega_i$. 
 
 In the following sections we derive the functions describing the indistinguishability and purity of photons for these sources after passing through two-photon, HOM-type, interferometers.

\section{On HOM Interference and Indistinguishability}

\begin{figure}[!t]
\centering
\includegraphics[scale=1]{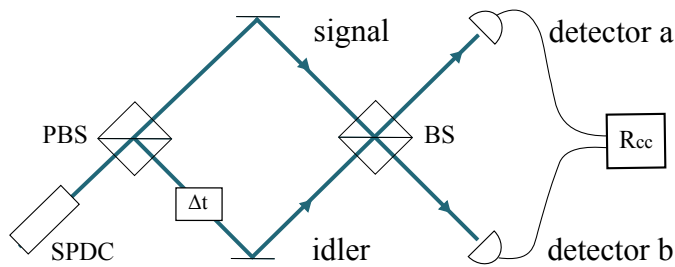}
\caption{A typical HOM interferometer, where a pair of photons generated, in this case in a Type-II  SPDC  process, are first spatially separated by a polarizing beam splitter (PBS) before being recombined on a beamsplitter (BS) to interfere. The coincidence probability $R_{cc}$ presents an interference pattern when measured as a function of the delay time $\Delta t$ between the photons arriving at the beamsplitter (BS).} \label{interferometer}
\end{figure}

The indistinguishability $I$ of two photons quantifies the overlap between their quantum states, i.e. for  states with no overlap, $I=0$, while for indistinguishable photons $I=1$. The most natural way to test this, is to try to make the photons bunch, which is exactly what Hong, Ou and Mandel did in their seminal paper about two-photon interference \cite{Hong87}. In a so-called HOM interferometer, as shown in figure  \ref{interferometer}, the signature interference dip can be observed if the detectors cannot resolve the  coherence time of the photons \cite{detectors}. If this holds, the probability $R_{cc}$ of having a coincidence between the outputs of the beamsplitter (BS) is given by the integral over the detection times ($\tau_a$, $\tau_b$) of the second order coherence function $G^{(2)}(\tau_a,\tau_b)= \langle   \rho \hat{E}_a^{(-)}(\tau_a) \hat{E}_b^{(-)}(\tau_b) \hat{E}_b^{(+)}(\tau_b) \hat{E}_a^{(+)}(\tau_a) \rangle$,  where $\hat{E}_{a}$ and $\hat{E}_{b}$ are the field operators at the detectors $a$ and $b$ \cite{Rubin94,Mandel95}, that is

\begin{eqnarray}
\label{RCCG2}
 R_{cc}&&=\int d\tau_a d\tau_b G^{(2)}(\tau_a,\tau_b).
 \end{eqnarray}
\noindent By calculating this probability as a function of the temporal delay $\Delta t$ between the photons, one can show that the coincidences have the  characteristic interference pattern given by
\begin{eqnarray}
\label{RCCi}
 R_{cc}(\Delta t)&&=\frac{1}{2}\left[1-I(\Delta t)\right],
\end{eqnarray}
where
\begin{eqnarray}
\label{IDt}
I(\Delta t)&=&\int  d\Omega_s d\Omega_i \Phi(\Omega_s,\Omega_i)\Phi^*(\Omega_i,\Omega_s) \exp{\left[-i (\Omega_s-\Omega_i)\Delta t\right]}.
\label{I}
\end{eqnarray}

\noindent The spectral indistinguishability between the signal and idler corresponds to the value of this function for $\Delta t=0$, $I=I(0)$  \cite{space}. Intuitively, the fact that the indistinguishability depends on the product of   $\Phi(\Omega_s,\Omega_i)$ and $\Phi^*(\Omega_i,\Omega_s)$ shows that $I(0)$ quantifies the effect of exchanging one photon with the other. 

In practice it is more convenient to extract the indistinguishability from the visibility of the interference pattern in \eref{RCCi}. In this case, the visibility is defined as the difference between the maximum and the minimum values of the interference pattern normalized to the maximum value, that is
\begin{eqnarray}
\label{Vdef}
V&=&\frac{\mathrm{max}[R_{cc}(\Delta t)]-\mathrm{min}[R_{cc}(\Delta t)]}{\mathrm{max}[R_{cc}(\Delta t)]}.
\end{eqnarray}

\noindent The maximum coincidence probability is obtained for a large $\Delta t$ and is equal to $\frac{1}{2}$, while the minimum occurs at $\Delta t=0$, and it is given by $\frac{1}{2}(1-I(0))$. So, for the interference pattern described by \eref{RCCi}, the visibility is equal to the indistinguishability: $V=I$.

\begin{figure}[thb]
\centering
\includegraphics[scale=0.50]{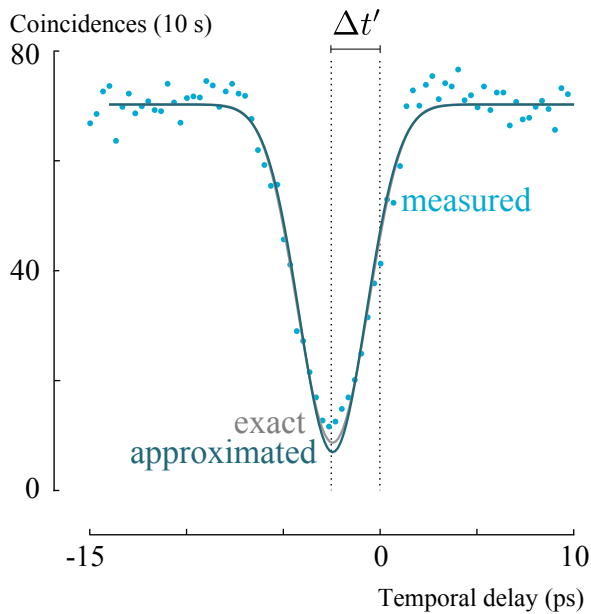}
\caption{The exact and approximated calculations for the HOM dip compared with a dip obtained experimentally.} \label{inds}
\end{figure}
If the state of the downconverted photons is described using \eref{aproxxmf}, the integral in \eref{IDt} has analytic solutions. In term of the parameters for the  SPDC  process, the indistinguishability  between the photons can now be simply given by
\begin{eqnarray}
\label{Iaprox}
I&=\left[\frac{4ab-4c^2}{(a+b)^2-4c^2}\right]^{1/2},
\end{eqnarray}

\noindent such that the photons are indistinguishable when $a=b$. The interference pattern obtained with the interferometer in figure  \ref{interferometer} is a dip  described by the function

\begin{eqnarray}
\label{RCCiaprox}
 R_{cc}(\Delta t)&&=\frac{1}{2}\left[1-I\exp\left[-2\frac{(\Delta t - \Delta t')^2}{a+b-2c}\right]\right], 
\end{eqnarray}

\noindent where both the width $I$ and the depth $a+b-2c$ of the dip  are  functions of the pump beam bandwidth, the filters, and the group velocity mismatch between the photons.  In non-degenerate configurations, where the group velocity mismatch is non-zero, $\Delta t'=L(N_s-N_i)/2$ is responsible for shifting the minimum of the dip.

To validate the approximated solution of \eref{RCCiaprox}, we compare it with the interference pattern  obtained by solving numerically the integrals of \eref{IDt} (using the exact mode function). We also compared this to experimental data obtained from a Type-II configuration where we generated orthogonally polarized photons from a $2$\,cm PPLN (periodically poled lithium niobate) \cite{Gayer08} crystal with a $780$\,nm pump laser beam with a $0.2$\,nm bandwidth. The photons are frequency degenerate at $1560$\,nm and  are filtered by identical $1.4$\,nm Gaussian filters. The results are illustrated in  figure \ref{inds}. The estimated visibility is $85.5\,\pm\,3.2\%$ for the experimental data,  $87.6\,\%$ for the exact calculation and $90.0\,\%$ when using the approximations.  The difference between the experimental data and the theoretical calculations can be explained by a small difference between the central frequency of the photons, due to fluctuations of the temperature of the crystal during the experiment. This difference is also responsible for the oscillations outside of the dip \cite{Ou88}.
\begin{figure}[htb]
\centering
\includegraphics[scale=0.50]{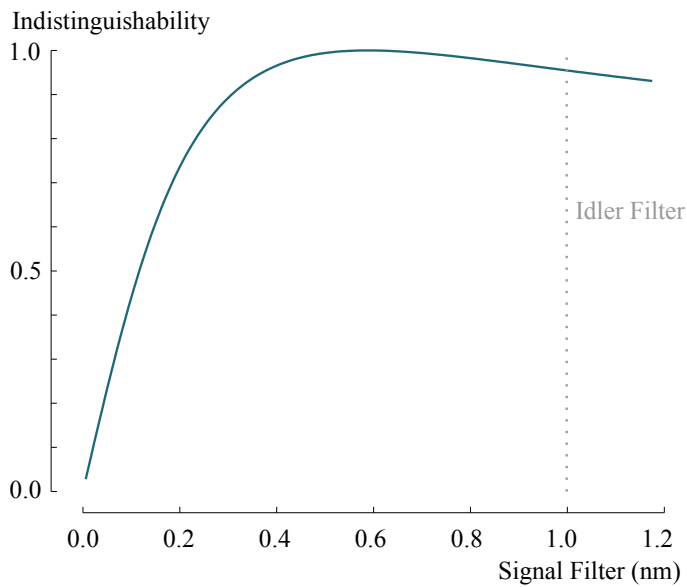}
\caption{Indistinguishability as a function of the bandwidth of the filter for the signal photon, when the idler is filtered with a $1$ nm filter.} \label{asymetric}
\end{figure} 

From \eref{abc}, one sees that perfect indistinguishability requires
\begin{eqnarray}
\frac{1}{\sigma_s^2}-\frac{1}{\sigma_i^2}=\alpha^2L^2(N_p-N_i)^2- \alpha^2L^2(N_p-N_s)^2,
\end{eqnarray}

\noindent and, since in Type-II configuration $Ns\neq N_i$, this condition cannot be satisfied using identical filters, unless they are significantly narrower than the natural bandwidth of the photons  \cite{ultranarrow}.  Therefore, even though it is rather counter-intuitive, to optimize the visibility of non-degenerate systems, one needs to use different filters for each photon. If we take the same experimental parameters as in figure \ref{inds}, then  figure  \ref{asymetric} shows how when the bandwidth of the idler filter is $1.0$ nm (FWHM) the maximum indistinguishability is obtained for a  $0.6$ nm filter for the signal. For identical filters, with $1.0$ nm bandwidth, the indistinguishability is only $94.0\%$. These values will obviously be different for the particular experimental scenario under consideration.

\section{On HOM interference and single photon Purity}

\begin{figure}[thb!]
\centering
\includegraphics[scale=0.8]{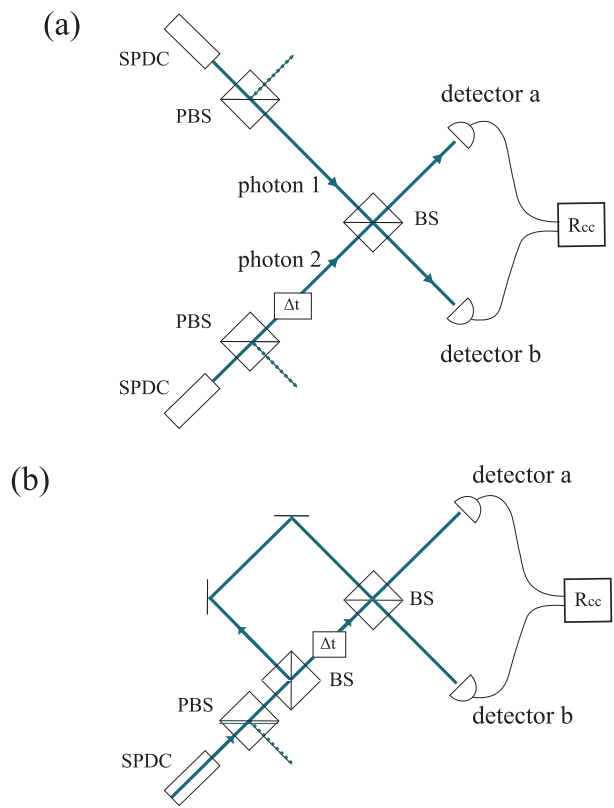}
\caption{Variations of the HOM schemes to measure the purity of one photon. The two copies of the photon, required for purity measurement, can be generated in two different but indistinguishable sources as in (a), or by two consecutive pulses of the pump laser as in (b). For this case the temporal distance between the pulses should be equal to the pulse-to-pulse time for the laser.} \label{singlephoton}
\end{figure}

While the indistinguishability is a property of one photon with respect to other, the purity is a property of a single photon state. The purity $P$ of a state $\rho$, is defined as the trace of its density matrix squared $P=Tr[\rho^2]$. Since it is a second order function of the density matrix, a measurement of the purity requires at least two copies of the state \cite{Brun04}. One of the most common ways to generate those copies is by using two identical crystals, c.f.  figure\,\ref{singlephoton}\,(a), where two photons, 1 and 2,  from independent SPDC sources are combined on a 50/50 BS.  The interference pattern obtained while varying the delay between their arrival times, will have unit visibility as a signature of perfect indistinguishability and purity \cite{Lee03,Mosley08,Mosley08a}.

Although numerous experiments have succeeded in interfering photons coming from independent sources, to obtain indistinguishable photons from different crystals is experimentally challenging. Variations on the input mode, the composition of the material or its temperature, the output couplings, among many other factors, will introduce distinguishability between the photons. We propose a way  to overcome this problem by generating the two photons in the same crystal by using, for instance, different pulses or different time bins. Figure \ref{singlephoton} (b) shows the principle whereby a temporal delay $\Delta t$ is set such that photons coming from consecutive pulses arrive simultaneously at different inputs of a 50/50 BS.   

Let $\rho_1$ and $\rho_2$ refer to photonic states from consecutive pump pulses 1 and 2, each of which is given by \eref{dms2}. The input state at the BS is given by the product state $\rho_1\otimes\rho_2$. The probability of measuring a coincidence as a function of the delay between the photons is
\begin{eqnarray}
\label{RCCp}
 R_{cc}(\Delta t) =\frac{1}{2}\left[1-P(\Delta t)\right],
\end{eqnarray}

\noindent provided there is one and only one photon in each of two consecutive pulses. The function $P(\Delta t)$ is given by

\begin{eqnarray}
\label{PDt}
P(\Delta t)&=&\int  d\Omega_1 d\Omega_2 \phi_1(\Omega_1,\Omega_2)\phi_2^*(\Omega_2,\Omega_1)\exp{\left[-i (\Omega_1-\Omega_2)\Delta t\right]}.
\end{eqnarray}

\noindent To simplify the notation we use the single-photon mode function given by \eref{mfs}. The visibility of the interference pattern described by \eref{RCCp} is $V=P(0)$. It is possible to show that $P(0)=Tr[\rho_1\rho_2]$, which can be also written as \cite{Lee03,Mosley08,Mosley08a}:

\begin{eqnarray}
\label{P}
V=Tr[\rho_1\rho_2]&=&\frac{Tr[\rho_1^2]+Tr[\rho_2^2]-2\|\rho_1-\rho_2\|^2}{2},\hspace{5mm}
\end{eqnarray}

\noindent where $\|\rho\|^2=Tr[\rho^{\dagger}\rho]$. Therefore, the visibility $V$ depends on both the indistinguishability between the two photons ($\|\rho_1-\rho_2\|^2$) and the purity of each of them ($Tr[\rho_1^2],Tr[\rho_2^2]$). If it is not possible to distinguish the cases where there is one photon in each pulse or, two photons in one of the pulses, the maximum visibility decreases to $V=P(0)/3$. A more general study on the effects of multi-pair emission on the interference patterns has previously been detailed elsewhere \cite{Cosme08, Wasilewski08, Poh09, Sekatski12}.

There are two trivial cases in which these quantities are independent. First, when both photons come from a maximally entangled pair, such that $Tr[\rho_1^2]=Tr[\rho_2^2]=0$, whereby the visibility $V$ is a measurement of the photon indistinguishability through the norm $\|\rho_1-\rho_2\|^2$.  Second, when the photons are totally indistinguishable  $\|\rho_1-\rho_2\|=0$, and then the visibility is a direct measurement of the purity of the single photon state $\rho_1$.  In our scheme, c.f. figure \ref{singlephoton} (b), it is reasonable to assume that an experiment is stable from pulse to pulse, such that successive photons are indistinguishable. Therefore, in this case, the visibility gives a direct estimation of the purity.

As before, the simplified mode function of \eref{aproxxmf} allows one to calculate analytically the interference pattern described by \eref{RCCp}.   In the case of indistinguishable input photons, this is thus given by
\begin{eqnarray}
\label{RCCpapprox}
 R_{cc}(\Delta t)&&=\frac{1}{2}\left[1-P\exp[-\frac{\Delta t^2}{a}]\right],
\end{eqnarray}

\noindent where the purity of the input state  $P=P(0)$, can be written as a function of the SPDC parameters from Eq.\,(\ref{abc}) as

\begin{eqnarray}
\label{Paprox}
P&=\sqrt{1-\frac{c^2}{ab}}.
\end{eqnarray}

\noindent $P = 0$ corresponds to the case where the generated photons are maximally mixed, which corresponds to $c^2 = ab$. The generation of pure photons is achieved for $ab>>c$, which is possible by using narrow filters, at the cost of reduced photon flux. Alternatively, one can set  $c=0$, i.e. from \eref{abc}:

\begin{eqnarray}
\frac{1}{\sigma_p^2}=-\alpha^2L^2(N_p-N_s)(N_p-N_i),
\end{eqnarray}

\noindent which is only achievable for certain materials \cite{Mosley08,Mosley08a,Eckstein11,Gerrits11,Svozilik11}.

\begin{figure}[htb]
\centering
\includegraphics[scale=0.50]{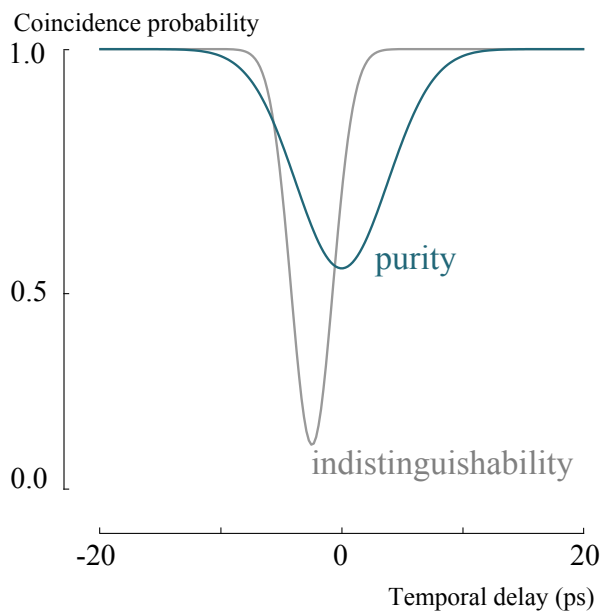}
\caption{Interference patterns in the coincidence probability for a two-photon state and for two independent photons. In the first case the visibility is equal to the indistinguishability of the photons and in the second case it is proportional to their purity.  We consider the same parameters as used for figure \,\ref{inds}, for which the purity is  $49.8\%$. } \label{pur}
\end{figure}

\section{Discussion}

One can see that there is a significant difference between the interference pattern obtained in a two-photon interferometer compared to the pattern obtained when two independent photons interfere, as described by \eref{RCCiaprox} and  \eref{RCCpapprox} and illustrated in figure \ref{pur}. These two equations are defined in terms of the SPDC parameters and hence provide a means to engineer the photonic states with the desired indistinguishability and purity. In this paper we have also proposed a novel interferometric  technique that allows for the purity to be measured using only a single SPDC source. This greatly simplifies the experimental complexity, and more importantly, does not rely on any assumption that photons from independent SPDC sources are indistinguishable. We expect that this simple and intuitive approach to characterizing and engineering SPDC photon sources will be of increasing importance for experiments combining multiple sources and engineering complex photonic networks.

\section*{Acknowledgements}
We would like to thank A. Valencia for fruitful discussions. This work was supported by the Swiss NCCR - Quantum Science and Technology and the EU project Q-essence.


\begin{thebibliography}{99}
\bibitem{detectors}The probability of having a coincidence is a convolution of the temporal response
of the detectors and the second order coherence function. If the detectors do not resolve the coherence time of the photons it is possible to consider their temporal response as flat. While,
 if the resolution of the detectors is smaller than the coherence time of the photons, the detectors act as temporal filters. In that case, the response of each detector also determine the temporal characteristics of the observed photons \cite{Sekatski12}.
\bibitem{space}For the full product state, including the spatial component, the total indistinguishability is a product of the indistinguishability in frequency given by  \eref{I} and the indistinguishability in space. If the mode function of the spatial part of the state is given by $\Phi_s({\bf q}_s,{\bf q}_i)$, the spatial indistinguishability  is given by 
$I_s=\int  d {\bf q}_s d{\bf q}_i \Phi_s({\bf q}_s,{\bf q}_i)\Phi_s^*({\bf q}_i,{\bf q}_s)
$, where ${\bf q}_s,{\bf q}_i$ are the transverse momentums.
\bibitem{ultranarrow} It is also always possible to generate indistinguishable photons by using identical ultra narrow filters, such that $1/\sigma_s>>\alpha L(N_p-N_s)$, $1/\sigma_i>>\alpha L(N_p-N_i)$ and $\sigma_s=\sigma_i$, but the number of photons will be drastically reduced.
\end{thebibliography}
\end{document}